# Can a nonradiating mode be externally excited? Nonscattering states vs. embedded eigenstates

Francesco Monticone,[1,*] Dimitrios Sounas,[2] Alex Krasnok,[3] and Andrea Alù [3*]

[1] Cornell University, School of Electrical and Computer Engineering, Ithaca, New York, 14853, USA

[2] Wayne State University, Department of Electrical and Computer Engineering, Detroit, MI 48202, USA

[3] Photonics Initiative, Advanced Science Research Center, City University of New York, NY 10031, USA

*Corresponding authors: francesco.monticone@cornell.edu ; aalu@gc.cuny.edu

*In this Letter, we discuss the general problem of exciting radiationless field distributions in open cavities, with the goal of clarifying recent findings on this topic. We point out that the radiationless scattering states, like anapoles, considered in several recent studies, are not eigenmodes of an open cavity; therefore, their external excitation is neither surprising nor challenging (similar to the excitation of nonzero internal fields in a transparent, or cloaked, object). Even more, the radiationless anapole field distribution cannot be sustained without the actual presence of external incident fields. Conversely, we prove that the Lorentz reciprocity theorem prevents the external excitation of radiationless optical eigenmodes, as in the case of embedded eigenstates and bound states in the continuum in open cavities. Our discussion clarifies the analogies and differences between invisible bodies, nonradiating sources, anapole scatterers and emitters, and embedded eigenstates, especially in relation to their external excitation.*

***Introduction*** – The ability to tailor optical scattering in anomalous and extreme ways, beyond what is achievable with conventional optical materials and structures, has been for several years one of the fundamental goals of optical metamaterials and nanophotonic systems [1]. Rapid progress in these fields has enabled the realization of a plethora of anomalous scattering effects, including invisibility [2–6], ultra-sharp Fano scattering resonances [7,8], non-scattering anapole scatterers [9–16], and bound states in the continuum or embedded eigenstates [17–22]. Scattering engineering plays a fundamental role in modern photonics research, for applications spanning from wavefront manipulation [23] and optical signal processing [24,25], to energy harvesting [26] and sensing [27], to mention just a few. Indeed, enhanced light-matter interactions and low-energy optical signal processing require the realization of highly confined fields in small open cavities (comparable or smaller than the wavelength), avoiding energy loss in the form of radiation or scattering, with the goal of increasing the lifetime of a highly-confined optical state and its interaction with matter. This opportunity has motivated and renewed interest in the concept of non-radiating sources [28], and has stimulated intense research efforts to realize optical scatterers supporting resonant field distributions that do not radiate. This growing area of research holds the potential to enable important advances in nanophotonics and quantum optics, for extreme light confinement in small open structures. Intriguingly, recent works (e.g., [9–16]) have discussed the possibility to observe and excite *radiationless anapole modes* supported by engineered scatterers. In particular, Ref. [12] suggested that the excitation of an anapole state "illustrates a case where the reciprocity condition is not violated, and yet, a radiationless mode can be excited by external illumination." Some of these studies, however, have introduced a degree of confusion regarding the nature of these radiationless optical states and the implications of reciprocity in these problems.

In this Letter, we discuss the general problem of exciting radiationless field distributions, and offer some considerations aimed at clarifying these relevant concepts and the relation among the different optical states involving reduced scattering and radiation. Most of the issues at stake stem from the use of the term "mode" in relation to the radiationless anapole, a terminology used in several recent papers that we feel it is prone to create confusion. In the following, we show that the so-called "anapole" is not a natural eigenmode of an open cavity formed by the scattering object. In other words, the anapole is not a self-sustained oscillation of the cavity satisfying boundary conditions on its own without an incident field. It is, instead, a resonant distribution of fields (or polarization currents) that can be excited by a suitable impinging field distribution [Fig. 1(a)]. It just so happens that this combination of induced polarization (and/or conduction) currents does not sustain a scattered wave, similar to other nonradiating induced field distributions extensively studied in the context of invisible or "cloaked" bodies (see, e.g., [2–6]). In contrast, we show that so-called "bound states in the continuum" or "embedded eigenstates", recently studied in several works [17–22], are actual examples of nonradiating eigenmodes of an open cavity, existing independently of the external field [Fig. 1(b)].

In general, the distinction between radiationless field distributions and radiationless eigenmodes is crucial when dealing with their excitation and reciprocity considerations.

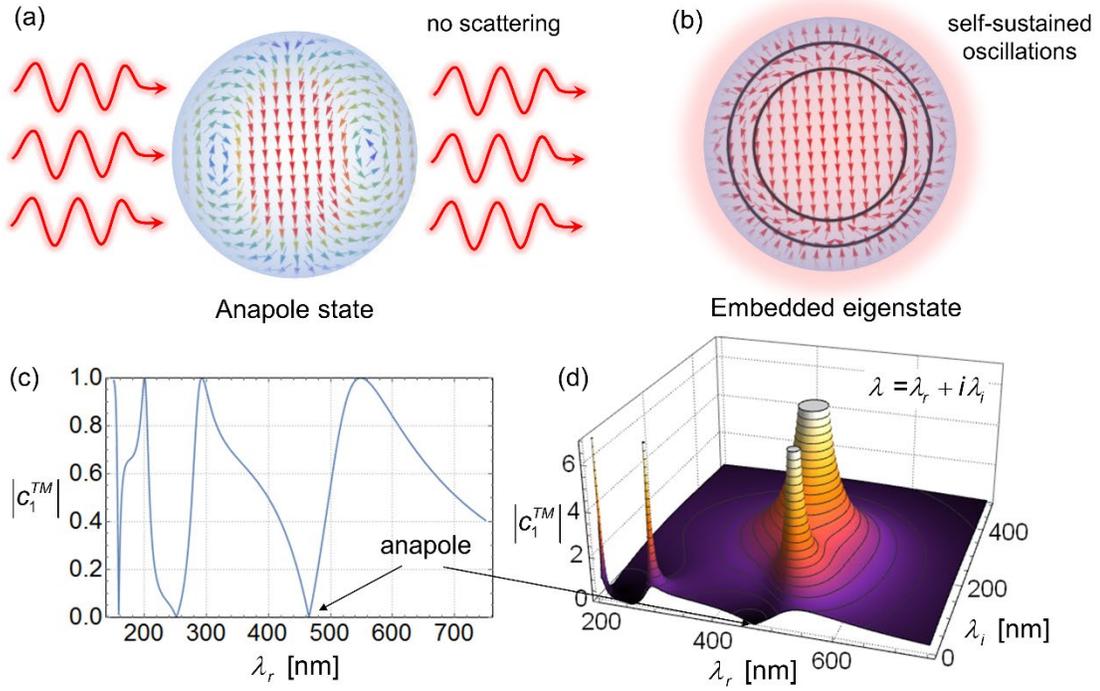

*Figure 1 – Illustrations of (a) an anapole state as a radiationless field distribution under an external illumination, and (b) an embedded eigenstate as a radiationless eigenmode of the system (self-sustained without an external field). (c) Amplitude of the electric dipolar Mie scattering coefficient on the real wavelength axis and (d) on the complex wavelength plane, for a dielectric sphere of radius a = 100 nm and n = 3.5. The longest-wavelength scattering zero corresponds to an anapole state. Scattering poles, corresponding to the eigenmodes of the open cavity, are confined to the upper-half of the complex wavelength plane ( $e^{-i\omega t}$ time-harmonic convention). The radiationless anapole is not an eigenmode of the open cavity, i.e., it cannot be self-sustained, and it can exist only in the presence of a suitable external excitation field.*

***Anapoles vs. Embedded Eigenstates*** – In the interest of simplicity, we focus our analysis on spherically-symmetric objects, for which the scattering problem can be rigorously solved analytically using Mie theory [29]. For a dielectric nanosphere with radius *a* = 100 nm and refractive index $n$ = 3.5, as considered in Ref. [12], we plot in Fig. 1(c) the Mie scattering coefficient of the first transverse-magnetic (TM) spherical harmonic, $c_1^{TM}$, which corresponds to the scattering contribution of the induced electric dipole, versus wavelength of excitation. The plot shows the typical alternation of scattering resonances, $\left|c_1^{TM}\right|=1$, and scattering zeros, $c_1^{TM}=0$, along the real wavelength axis. When the scatterer is illuminated by a generic field distribution containing the first TM harmonic (for example, a plane wave), the polarization currents induced in the sphere will sustain scattered fields whose electric-dipolar contribution is weighed by $c_1^{TM}$ according to Fig. 1(c), as a function of the impinging frequency.

In particular, if the object is excited at the frequency of a scattering zero of $c_1^{TM}$, the induced polarization density distribution will be set up in a way that does not produce any net electric dipole moment. In other words, the scatterer is invisible for that specific spherical harmonic, even though the object is polarized by the excitation and the internal fields are non-zero, namely, the incident field induces a radiationless distribution of polarization fields and/or currents. For the lowest-frequency scattering zero in Fig. 1(c), the scattering is dominated by the electric dipole contribution, and therefore the total scattering from the object can be made small at this frequency, which corresponds to the radiationless anapole condition. While this nonradiating polarization distribution can indeed be interpreted as arising from the cancellation of quasi-static Cartesian dipolar and toroidal contributions, as discussed in Ref. [9,12], we would like to stress that *it is not different from any other scattering zero of the dipolar Mie coefficient*, as observed in Fig. 1(c) along the wavelength axis, other than the fact that, given the long wavelength, all other multipolar scattering contributions happen to be small. In addition, while the zeros in Fig. 1(c) are inherent to the scattering response of a uniform dielectric sphere, zeros at arbitrary frequency positions can be induced by covering the object with suitably tailored shells (for example, a plasmonic layer), as conventionally done in cloaking by scattering cancellation [3,6]. Indeed, the radiationless anapole is physically and phenomenologically consistent with the large body of work on cloaked or low-scattering objects.

At this point, it is important to recognize that nonradiating scattering states, as in the case of anapoles and some cloaked states (e.g., [8]), can produce very large internal fields, while at the same time the external scattering is suppressed. In fact, the absence of radiation loss may partially contribute to enhancing the stored energy associated with these fields. For this reason, nonradiating scattering states with large enough fields may indeed look like resonant modes, and therefore be confused with radiationless eigenstates of the open cavity. What is important to remark, however, is that the anapole or other cloaked states are not at all eigenmodes of the scattering structure. Eigenmodes are self-sustained field distributions that correspond to the poles of the Mie scattering coefficients. Due to the presence of radiation loss, these poles are confined to the upper-half of the complex-frequency plane, as shown in Fig. 1(d) for the dipolar coefficient $c_1^{TM}$ for the same dielectric sphere considered above. On the contrary, in order to exist, the anapole state requires the presence of an external excitation, since its field distribution cannot satisfy the boundary-value problem without the presence of external fields (more on this point below). In light of these considerations, and the calculations in Fig. 1(d), it is clear that *the anapole state is not an eigenmode, and no self-sustained modal distribution of the dielectric sphere is radiationless (no pole sits on the real frequency axis)*.

In this context, it is relevant to highlight the analogies between the anapole state (scattering zero) in 2D/3D geometries and the common *Fabry-Perot tunneling effect* in a 1D dielectric slab resonator [30]. Indeed, at the Fabry-Perot zero, an incident wave can induce strong fields in the slab that do not perturb the impinging field distribution, offering zero reflections and full transmission. Similar to the anapole, this effect is also caused by interference between different partial waves (co- and counter-propagating waves), which do not produce net back radiation. Indeed, for a dielectric object both the Mie scattering coefficients in 2D/3D and the reflection coefficient in 1D exhibit an alternation of zeros and peaks along the real frequency axis. Importantly, in both cases, the internal fields are not necessarily minimized at the zeros.

Having established the difference between a nonradiating induced current distribution and a nonradiating eigenmode, it is indeed important to explore the possibility of exciting them by an external field, and the associated conditions imposed by reciprocity. In light of the above discussion, it is evident that *being able to excite a radiationless field distribution like the anapole is not surprising*: it simply requires exciting an object at the frequency of a scattering zero, as typically done in the context of cloaking devices [6]. In general, special excitations are not at all necessary to excite such radiationless field distributions, as long as the excitation contains (i.e., it is not orthogonal to) the relevant harmonic (a simple plane wave, which contains all spherical harmonics, would be sufficient to excite an anapolar field distribution). However, a suitably tailored excitation, like the one in Ref. [12], targeting the excitation of the specific spherical harmonic of interest, may allow suppressing higher-order scattering contributions, which can be non-negligible at the frequency of the anapole when the size of the object is not too small compared to the wavelength.

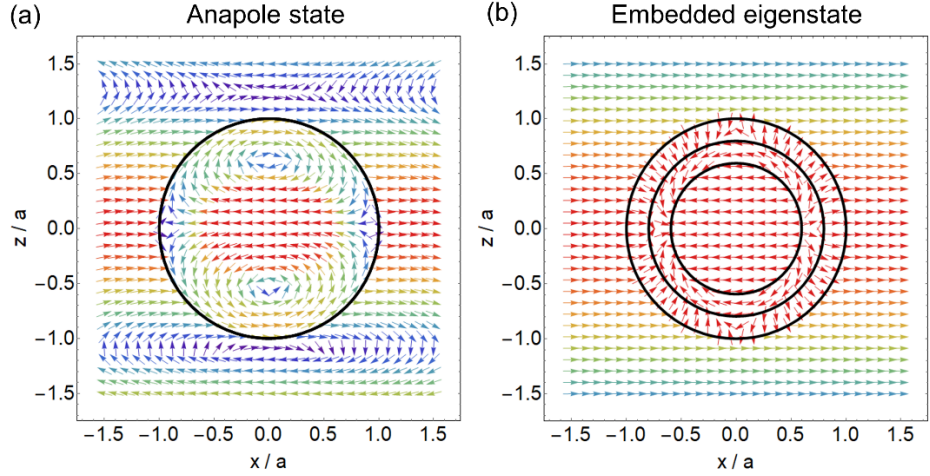

*Figure 2 – Electric field vector distribution (time-snapshots) for (a) a dielectric sphere supporting a radiationless anapolar field distribution, as in [12], and (b) the layered metallo-dielectric sphere considered in Ref. [20], supporting an embedded scattering eigenstate. Warmer colors correspond to higher fields. The scatterers are illuminated by a plane wave propagating toward the +z-axis at the relevant frequency. The fields are plotted on the central plane of the sphere. Dimensions are normalized to the outer radius of the scatterers.*

The presence of external incident fields is actually *necessary* to set up the nonradiating anapole distribution, which, as pointed out above, does not satisfy, on its own, the continuity of the tangential fields on the sphere boundary. To further clarify this issue, we plot in Fig. 2(a) the electric-field vector distribution for a dielectric sphere illuminated by a plane wave, at the frequency of the radiationless anapole condition, on the central plane of the sphere. Not surprisingly, the incident wave is not perturbed by the presence of the scatterer, since we operate at a zero of the dominant dipolar scattering coefficient. Most importantly, by inspecting the field distribution on the surface of the sphere, it is evident that the internal anapolar fields would not be able to satisfy the condition of continuous tangential fields on the surface of the sphere, if the external field were zero! The incident field is absolutely necessary to satisfy this boundary condition. Indeed, as shown in our transient simulations reported in the following, an anapolar field distribution starts radiating as soon as the incident wave is switched off at a certain time instant, to compensate for the lack of excitation fields in the outside region. Particularly revealing is the comparison with an actual nonradiating eigenmode, for example the embedded eigenstate supported by the metallo-dielectric layered spheres considered in Ref. [20]. The corresponding electric-field vector distribution is shown in Fig. 2(b) at the frequency of the scattering zero near an embedded eigenstate. In this case, the internal electric field is almost purely radial on the surface of the spherical scatterer. Hence, the continuity condition of the tangential field components is satisfied even if the external fields are zero (also the continuity of the normal component of the displacement field is satisfied, even with zero external fields, $\varepsilon_1 \hat{n} \cdot \mathbf{E}_1 = \varepsilon_2 \hat{n} \cdot \mathbf{E}_2 = 0$, because the permittivity of the outer shell is required to be zero at the embedded-eigenstate condition). Hence, unlike the anapole field distribution, an embedded eigenstate is indeed a nonradiating eigenmode of the system, existing independently of the external field. In addition, we stress that, although in Fig. 2(b) we have chosen to work at the frequency of a scattering zero, which is always present around a nearly-ideal embedded eigenstate, the embedded eigenstate itself does not imply only zero scattering, but the combined presence of a scattering pole and zero, which collide and cancel out at a real frequency in the ideal case [20].

***Role of Lorentz Reciprocity*** – To understand the role of electromagnetic reciprocity in the excitation of radiationless field/current distributions, consider the Lorentz reciprocity theorem applied to the general case in Fig. 3:

$$\int_{V_s} \mathbf{J}_1 \cdot \mathbf{E}_2 \, dV = \int_V \mathbf{J}_2 \cdot \mathbf{E}_1 \, dV, \tag{1}$$

where $\mathbf{J}_1$ is an electric current in a finite volume $V_s$ that produces the electric field $\mathbf{E}_1$ in the external volume $V \setminus V_s$, whereas the electric current $\mathbf{J}_2$ in the external volume $V \setminus V_s$ produces the field $\mathbf{E}_2$ inside the volume $V_s$ (all fields are defined in free space). If the current distribution $\mathbf{J}_1$ is nonradiating, then $\mathbf{E}_1 = 0$ in $V \setminus V_s$, and Eq. (1) reduces to

$$\int_{V_s} \mathbf{J}_1 \cdot \mathbf{E}_2 \, dV = 0. \tag{2}$$

Eq. (2) implies that any continuous non-radiating current distribution in a finite volume $V_s$ is orthogonal to any field solution in $V_s$ that is consistent with Maxwell's equations. This is a particularly strong result, which confirms and generalizes the findings by E. Wolf in the context of nonradiating sources (see Eq. 3.6 in Ref. [28]). Eq. (2) may be applied to a scatterer with an internal non-radiating field distribution, $\mathbf{E}_{int}$, if $\mathbf{J}_1$ is taken as the induced polarization current, i.e., $\mathbf{J}_1 = -i\omega(\varepsilon - \varepsilon_0)\mathbf{E}_{int}$, with $\varepsilon$ the permittivity of the scatterer, and $\mathbf{E}_2$ is taken as the incident field $\mathbf{E}_{inc}$ in the absence of the scatterer. Hence, we can write:

$$\int_{V_s} (\varepsilon(\mathbf{r}) - \varepsilon_0) \mathbf{E}_{int} \cdot \mathbf{E}_{inc} \, dV = 0. \tag{3}$$

As correctly pointed out in Ref. [12], if a certain field distribution $\mathbf{E}_{int}$ in a volume $V_s$ does not radiate, Lorentz reciprocity only requires the integral in Eq. (2) to identically vanish, whereas it does not imply that the nonradiating field distribution is zero everywhere inside the scatterer (in other words, it does not require the object to be an impenetrable system, such as a perfectly conducting cavity). This is consistent with the recent literature on cloaked objects, which support induced fields in the cloak and/or the object, but no scattering in the outside region. The situation is very different if the nonradiating field distribution $\mathbf{E}_{int}$ corresponds to an eigenmode of the system, as in the case of embedded eigenstates.

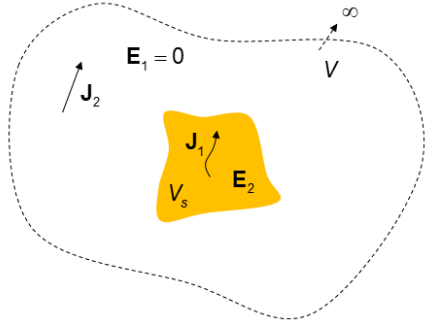

*Figure 3 – Geometry for the application of the Lorentz reciprocity theorem to a nonradiating current distribution $\mathbf{J}_1$. V indicates the entire (unbounded) volume.*

To clarify the behavior of an ideal nonradiating eigenmode under a causal external excitation, and its implications in terms of reciprocity, it is preferable not to rely on a steady-state frequency-domain analysis, because steady state may never be reached for an eigenmode with infinite Q factor. Instead, a time-domain analysis avoids any confusion regarding steady-state operation. In Supporting Information, we show that, for a given temporal excitation, the field distribution inside a spherically-symmetric cavity can be expanded into a set of eigenmodes [31–38]

$$\mathbf{E}_s(\mathbf{r},t) = \sum_n a_n(t) \mathbf{E}_n(\mathbf{r}), \qquad (4)$$

where $a_n$ are unknown coefficients that indicate how strongly a certain mode is excited by the impinging field distribution. The eigenvectors $\mathbf{E}_n$ correspond to the eigenmodes of the scatterer, which can be shown to form a complete set *inside* canonical geometries, for example spherically-symmetric material bodies [31–38]. The eigenfrequencies $\omega_n$ are generally complex due to the open nature of the considered system, and they correspond to the complex poles in Fig. 1(d). Embedded eigenstates represent a relevant exception, as their eigenfrequency $\omega_n$ is purely real despite the system being open to radiation.

Without loss of generality, we choose a plane-wave square pulse of duration $T_p$ and central frequency $\omega$, as an example of a causal incident field impinging on the scatterer. Intuitively, we expect that the incident field may excite the complex eigenstates of the scatterer as it impinges on it. Then, when the incident field leaves the object, the modal amplitude is expected to decay exponentially as the excited modes release the stored energy like harmonic oscillators damped by radiation loss. In Supporting Information, we show that, in the regime after the incident field has completely left the scatterer, the modal expansion coefficients are given by

$$a_n(t) = \frac{i}{2} \frac{e^{-i(\omega-\omega_n)T_p} - 1}{(\omega-\omega_n)\langle \mathbf{E}_n | \mathbf{E}_n \rangle} e^{-i\omega_n t}$$
$$\times \int_{V_s} (\varepsilon(\mathbf{r}') - \varepsilon_0) \mathbf{E}_n(\mathbf{r}') \cdot \mathbf{E}_{inc}(\mathbf{r}', \omega_n) d\mathbf{r}' \cdot \qquad (5)$$

If we now assume that the *n*-th eigenmode of the system is non-radiating, namely, it is an embedded eigenstate, two things should be recognized: first, $\omega_n$ becomes real, hence the real excitation frequency $\omega$ may be chosen to be exactly equal to $\omega_n$; second, the spatial integral in (5), namely, the integral of a nonradiating polarization current times the incident field, becomes equal to the integral in (3), and it is therefore zero. This implies that the modal amplitude vanishes, $a_n = 0$, for any excitation frequency $\omega$, including for $\omega \to \omega_n$ [in this limit, the prefactor in (5) becomes equal to $T_p e^{-i\omega_n t}/2$, which is finite for finite pulse duration].

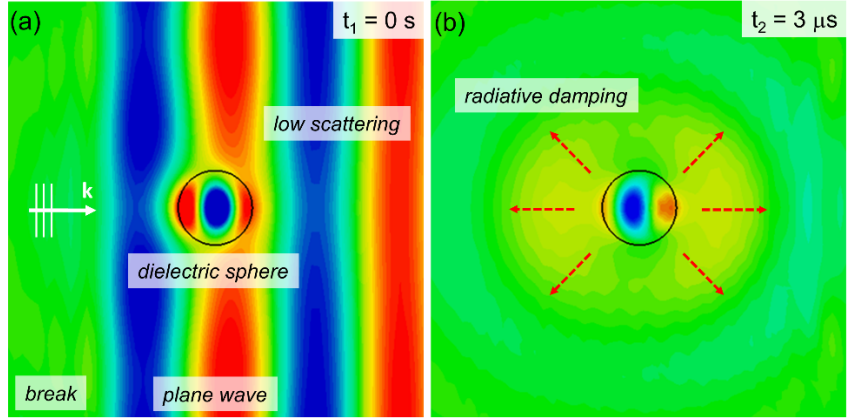

*Figure 4 – Time-snapshots of the (out-of-plane) electric field distribution around a dielectric sphere supporting an anapole state (same scatterer considered in Figs. 1,2). (a) Transient plane-wave excitation with central frequency tuned at the anapole scattering zero in Fig. 1(c,d). (b) Radiative damping of the anapole state after the incident field has left the scatterer. A time-domain animation of the field distribution is available as Supplementary Material.*

These considerations indicate that, if a nonradiating field distribution corresponds exactly to an eigenmode of the scatterer (i.e., it is an embedded eigenstate), then the energy stored in the nonradiating eigenmode under an external causal illumination must be identically zero! If, instead, the eigenmode is not perfectly radiationless, the spatial integral in (5) is not exactly zero, and the modal amplitude evolves in time as $e^{-i\omega_n t}$, decaying exponentially according to the imaginary part of the eigenfrequency, as expected, which indicates that the resonant mode can be excited and the energy stored in the mode is non-zero.

As a relevant example, Fig. 4 shows the transient behavior for the same dielectric sphere as in Fig. 1 excited at its anapole frequency. The sphere is illuminated by a plane-wave pulse, whose wavefront intensity abruptly drops to zero, as shown in the time-snapshot of the field distribution in Fig. 4(a) ($t_1 = 0$). As long as the incident field impinges on the sphere, the scattering is low, consistent with the dipolar scattering zero in Fig. 1. However, after the wave has completely left the scatterer, the anapole field distribution – which cannot satisfy the boundary conditions on the surface of the sphere on its own – starts radiating and rapidly decays in time. In about 2 μs, approximately half of the stored energy is released, as seen in Fig. 4(b). This result confirms that the anapole state is nonradiative only in steady state, when a non-zero monochromatic incident field is present, whereas it rapidly decays due to radiation when the excitation vanishes. Moreover, in the case of an anapolar field distribution, which is not an eigenmode of the system, but is established by the interaction of different radiation-damped complex eigenmodes in the framework of Eq. (4) (like Fano resonances [39]), the scatterer releases the stored energy as the superposition of multiple exponentially decaying oscillations at different frequencies.

Finally, we would like to briefly mention the interesting proposal of *anapole-based nanolasers* [15]. Both semi-classical and quantum laser theories [40–43] require a system to have a well-defined mode with a pole (of the relevant scattering coefficient or S-matrix eigenvalue) in the complex plane. Exact (radiation and absorption) loss compensation due to the introduction of gain may move this pole onto the real frequency axis, which makes the system enter a self-sustained lasing regime [40,41]. However, due to the nature of the anapole state as a zero, not a pole, of the scattering coefficient, an anapole state cannot be used, directly, to achieve lasing, whereas we expect the pole constellation around the anapole [as in Fig. 1(d)] to be able to contribute to amplification and lasing if optical gain is suitably introduced in the resonant scatterer.

***Conclusion*** – In summary, we have discussed the similarities and differences between two classes of radiationless scattering states, anapoles and embedded eigenstates, focusing in particular on the issue of exciting these field distributions in an open cavity and the implications of reciprocity. We have demonstrated that, in the case of an ideal embedded eigenstate, the radiationless mode cannot be excited at all by external sources, and the incident field, while entering the sphere, cannot induce the eigenstate fields, as a result of reciprocity. In other words, the polarization fields induced at the frequency of an embedded eigenstate are orthogonal to the embedded eigenstate fields, which, in a linear reciprocal system, cannot couple to external excitation. However, *there is no limit* to how close one can approach this nonradiating state in the lossless limit [17–22], implying that highly resonant fields may be excited in the strong proximity of the ideal embedded eigenstate condition. On the contrary, we have shown that the anapole is not an eigenmode of the scattering system; instead, it is an induced radiationless field distribution corresponding to a zero of the dipolar scattering coefficient. The anapole state is therefore nonradiative only in the presence of an external monochromatic incident field, and it decays into radiation when the excitation is switched-off. Hence, the external excitation of the anapole state is neither surprising nor challenging.

Given the generality of the discussed concepts of linear scattering theory and reciprocity, similar considerations on radiationless states directly apply to different realms of wave scattering physics, from electromagnetic and acoustic waves, to elastic and matter waves.

**Acknowledgements**

F.M. acknowledges support from the Air Force Office of Scientific Research (AFOSR) (FA9550-19-1-0043). A.A. acknowledges support from AFOSR, the National Science Foundation and the Simons Foundation.

**References**

1. F. Monticone and A. Alù, "Metamaterial, plasmonic and nanophotonic devices," Reports Prog. Phys. **80**, 036401 (2017).

2. M. Kerker, "Invisible bodies," J. Opt. Soc. Am. **65**, 376–379 (1975).

3. A. Alù and N. Engheta, "Achieving transparency with plasmonic and metamaterial coatings," Phys. Rev. E **72**, 16623 (2005).

4. U. Leonhardt, "Optical Conformal Mapping," Science **312**, 1777–1780 (2006).

5. J. B. Pendry, D. Schurig, and D. R. Smith, "Controlling electromagnetic fields," Science **312**, 1780–2 (2006).

6. R. Fleury, F. Monticone, and A. Alù, "Invisibility and cloaking: Origins, present, and future perspectives," Phys. Rev. Appl. **4**, (2015).

7. A. E. Miroshnichenko, S. Flach, and Y. S. Kivshar, "Fano resonances in nanoscale structures," Rev. Mod. Phys. **82**, 2257–2298 (2010).

8. F. Monticone, C. Argyropoulos, and A. Alù, "Multilayered Plasmonic Covers for Comblike Scattering Response and Optical Tagging," Phys. Rev. Lett. **110**, 113901 (2013).

9. K. V. Baryshnikova, D. A. Smirnova, B. S. Luk'yanchuk, and Y. S. Kivshar, "Optical Anapoles: Concepts and Applications," Adv. Opt. Mater. **1801350**, 1801350 (2019).

10. A. E. Miroshnichenko, A. B. Evlyukhin, Y. F. Yu, R. M. Bakker, A. Chipouline, A. I. Kuznetsov, B. Luk'yanchuk, B. N. Chichkov, and Y. S. Kivshar, "Nonradiating anapole modes in dielectric nanoparticles," Nat. Commun. **6**, 8069 (2015).

11. Y. Yang, V. A. Zenin, and S. I. Bozhevolnyi, "Anapole-Assisted Strong Field Enhancement in Individual All-Dielectric Nanostructures," ACS Photonics **5**, 1960-1966 (2018).

12. L. Wei, Z. Xi, N. Bhattacharya, and H. P. Urbach, "Excitation of the radiationless anapole mode," Optica **3**, 799 (2016).

13. A. K. Ospanova, I. V. Stenishchev, and A. A. Basharin, "Anapole Mode Sustaining Silicon Metamaterials in Visible Spectral Range," Laser Photon. Rev. **12**, 1800005 (2018).

14. B. Luk'yanchuk, R. Paniagua-Domínguez, A. I. Kuznetsov, A. E. Miroshnichenko, and Y. S. Kivshar, "Suppression of scattering for small dielectric particles: anapole mode and invisibility," Philos. Trans. R. Soc. A Math. Phys. Eng. Sci. **375**, 20160069 (2017).

15. J. S. Totero Gongora, A. E. Miroshnichenko, Y. S. Kivshar, and A. Fratalocchi, "Anapole nanolasers for mode-locking and ultrafast pulse generation," Nat. Commun. **8**, 15535 (2017).

16. K. Koshelev, G. Favraud, A. Bogdanov, Y. Kivshar, and A. Fratalocchi, "Nonradiating photonics with resonant dielectric nanostructures," Nanophotonics (2019).

17. C. W. Hsu, B. Zhen, A. D. Stone, J. D. Joannopoulos, and M. Soljačić, "Bound states in the continuum," Nat. Rev. Mater. **1**, 16048 (2016).

18. C. W. Hsu, B. Zhen, J. Lee, S.-L. Chua, S. G. Johnson, J. D. Joannopoulos, and M. Soljačić, "Observation of trapped light within the radiation continuum.," Nature **499**, 188–91 (2013).

19. M. G. Silveirinha, "Trapping light in open plasmonic nanostructures," Phys. Rev. A **89**, 023813 (2014).

20. F. Monticone and A. Alù, "Embedded Photonic Eigenvalues in 3D Nanostructures," Phys. Rev. Lett. **112**, 213903 (2014).

21. F. Monticone, H. M. Doeleman, W. Den Hollander, A. F. Koenderink, and A. Alù, "Trapping Light in Plain Sight: Embedded Photonic Eigenstates in Zero-Index Metamaterials," Laser Photon. Rev. **12**, 1700220 (2018).

22. H. M. Doeleman, F. Monticone, W. Den Hollander, A. Alù, and A. F. Koenderink, "Experimental observation of a polarization vortex at an optical bound state in the continuum," Nat. Photonics **12**, 397–401 (2018).

23. N. Yu and F. Capasso, "Flat optics with designer metasurfaces," Nat. Mater. **13**, 139–150 (2014).

24. A. Silva, F. Monticone, G. Castaldi, V. Galdi, A. Alù, and N. Engheta, "Performing mathematical operations with metamaterials.," Science **343**, 160–3 (2014).

25. H. Kwon, D. Sounas, A. Cordaro, A. Polman, and A. Alù, "Nonlocal Metasurfaces for Optical Signal Processing," Phys. Rev. Lett. **121**, 173004 (2018).

26. J. a Schuller, E. S. Barnard, W. Cai, Y. C. Jun, J. S. White, and M. L. Brongersma, "Plasmonics for extreme light concentration and manipulation.," Nat. Mater. **9**, 193–204 (2010).

27. M. E. Stewart, C. R. Anderton, L. B. Thompson, J. Maria, S. K. Gray, J. A. Rogers, and R.

G. Nuzzo, "Nanostructured Plasmonic Sensors," Chem. Rev. **108**, 494–521 (2008).

28. K. Kim and E. Wolf, "Non-radiating monochromatic sources and their fields," Opt. Commun. **59**, 1–6 (1986).

29. D. R. Bohren, Craig F, Huffman, *Absorption and Scattering of Light by Small Particles* (Wiley-VCH Verlag GmbH, 1998).

30. A. Kavokin, J. J. Baumberg, G. Malpuech, and F. P. Laussy, *Microcavities* (Oxford University Press, 2007).

31. P. T. Kristensen, C. Van Vlack, and S. Hughes, "Generalized effective mode volume for leaky optical cavities," Opt. Lett. **37**, 1649 (2012).

32. P. T. Kristensen and S. Hughes, "Modes and Mode Volumes of Leaky Optical Cavities and Plasmonic Nanoresonators," ACS Photonics **1**, 2–10 (2014).

33. K. M. Lee, P. T. Leung, and K. M. Pang, "Dyadic formulation of morphology-dependent resonances I Completeness relation," J. Opt. Soc. Am. B **16**, 1409 (1999).

34. C. Sauvan, J. P. Hugonin, I. S. Maksymov, and P. Lalanne, "Theory of the Spontaneous Optical Emission of Nanosize Photonic and Plasmon Resonators," Phys. Rev. Lett. **110**, 237401 (2013).

35. R.-C. Ge, P. T. Kristensen, J. F. Young, and S. Hughes, "Quasinormal mode approach to modelling light-emission and propagation in nanoplasmonics," New J. Phys. **16**, 113048 (2014).

36. P. T. Leung, S. Y. Liu, and K. Young, "Completeness and orthogonality of quasinormal modes in leaky optical cavities," Phys. Rev. A **49**, 3057–3067 (1994).

37. P. Lalanne, W. Yan, K. Vynck, C. Sauvan, and J.-P. Hugonin, "Light Interaction with Photonic and Plasmonic Resonances," Laser Photon. Rev. **12**, 1700113 (2018).

38. D. A. Powell, "Interference between the Modes of an All-Dielectric Meta-atom," Phys. Rev. Appl. **7**, 034006 (2017).

39. B. Stout and R. McPhedran, "Egocentric physics: Just about Mie," EPL (Europhysics Lett. **119**, 44002 (2017).

40. H. Haus, *Waves and Fields in Optoelectronics* (Prentice Hall, 1984).

41. L. Ge, Y. D. Chong, and A. D. Stone, "Steady-state ab initio laser theory: Generalizations and analytic results," Phys. Rev. A **82**, 063824 (2010).

42. G. Björk, A. Karlsson, and Y. Yamamoto, "Definition of a laser threshold," Phys. Rev. A **50**, 1675–1680 (1994).

43. L. Reeves, Y. Wang, and T. F. Krauss, "2D Material Microcavity Light Emitters: To Lase or Not to Lase?," Adv. Opt. Mater. **6**, 1800272 (2018).